\documentclass [11pt,useAMS] {article}
\usepackage{amsmath}   
\usepackage{graphicx}   
\usepackage{verbatim}   
\usepackage{color} 
\usepackage{float}     
\usepackage{subfigure} 
\usepackage{hyperref}
\usepackage{latexsym}
\usepackage{caption}
\usepackage{wrapfig}
\usepackage{soul}
\usepackage[margin=0.6in]{geometry}

\begin{document}

\noindent \textbf{The Relation Between Fundamental Constants and Particle Physics Parameters}\\
Proceedings of the Conference on Varying Constants and Fundamental Cosmology VARCOSMOFUN16\\
\\
Rodger I. Thompson$^{1}$

$^{1}$Steward Observatory and Department of Astronomy, University of Arizona: 
 rit@email.arizona.edu

\section{Abstract}
The observed constraints on the variability of the proton to electron mass ratio $\mu$ and the fine
structure constant $\alpha$ are used to establish constraints on the variability of the Quantum Chromodynamic
Scale and a combination of the Higgs Vacuum Expectation Value and the Yukawa couplings.  Further model 
dependent assumptions provide constraints on the Higgs VEV and the Yukawa couplings separately.   A primary 
conclusion is that limits on the variability of dimensionless fundamental constants such as $\mu$ and $\alpha$
provide important constraints on the parameter space of new physics and cosmologies.


\section{Introduction}
Over the past two decades there has been a renewed interest in measuring fundamental dimensionless constants such as
the proton to electron mass ratio $\mu$ and the fine structure constant $\alpha$ in the early universe.  Modulo some
possibilities discussed at this conference, impressive constraints on the variation of $\mu$ and $\alpha$ have been
established over time periods that span a significant fraction of the age of the universe.  Here the implications of
those constraints are examined in terms of the stability of three basic physics parameters, the Quantum Chromodynamic
Scale $\Lambda_{QCD}$, the Higgs Vacuum Expectation Value $\nu$ and the Yukawa Couplings $h$.  Previous
reports of a possible variation of $\alpha$ \cite{web01} spurred significant efforts to account for the variation in terms of
varying $\Lambda_{QCD}$, $\nu$ and $h$ \cite{cam95, cal02,lan02,lan04,din03,cha07,coc07,den08,uza11}.  These
efforts form the basis of the present work except that the process is reversed in that constraints on the variability
of the physics parameters is established in terms of the limits on the variability of $\mu$ and $\alpha$.  Most
of the earlier efforts limited the variabililty to only one of the physics parameters, usually $\Lambda_{QCD}$.  In
contrast the work of \cite{coc07} considered the possibility of all three parameters varying.  As such it is
the primary reference in this work.

\section{Observational Constraints} \label{s-obscon}
There are relatively few constraints on the variation of $\mu$ since molecular spectra provide the primary limits
on the variability of $\mu$ \cite{thm75} and only a few molecular spectra at high redshift exist.  On the other hand
radio observations of molecular spectra at moderate redshifts provide a significantly tighter constraint on $\mu$ than
exists for $\alpha$.  The current limits on both $\mu$ and $\alpha$ are given below.

\subsection{$\mu$ constraints} \label{ss-mucon}
\begin{figure}[H]
\centering
\includegraphics[width=10cm]{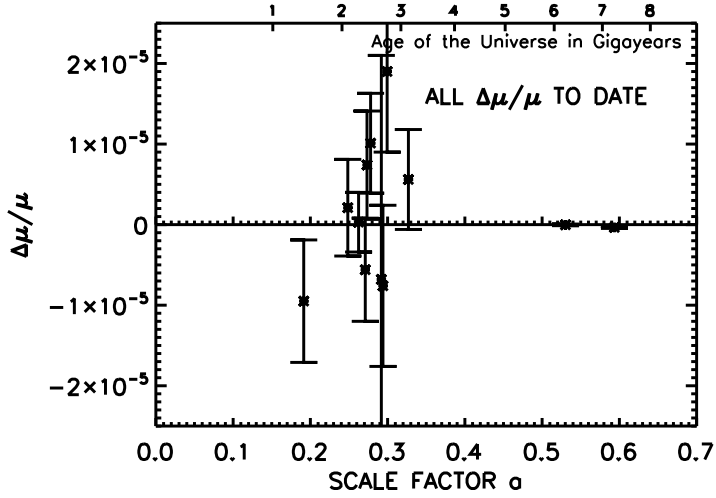}
\caption{All of the observational constraints on $\Delta \mu / \mu$ from radio ($z < 1$)
and optical ($z > 1$) observations  plotted versus the scale factor $a = 1/(1+z)$.  All constraints are 
at the $1 \sigma$ level. The low redshift radio constraints are difficult to see at the scale of this plot.
The age of the universe in gigayears is plotted on the top
axis and in fig.~\ref{fig-rad}.}  \label{fig-obs}
\end{figure}

\begin{figure}[H]
\centering
\includegraphics[width=10cm]{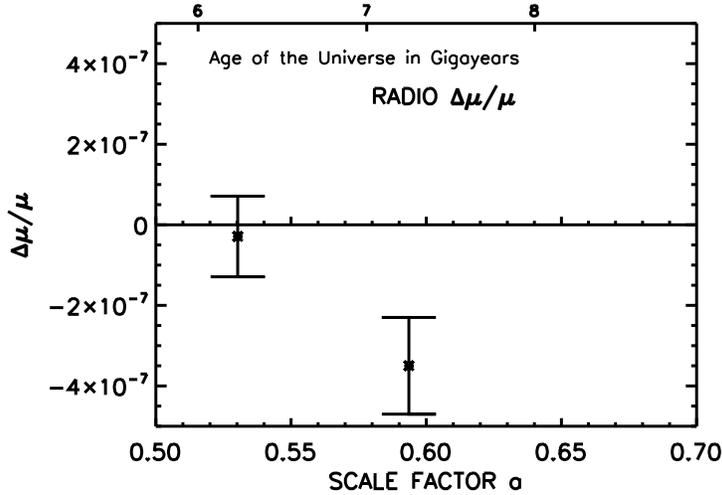}
\caption{The low redshift radio $\Delta \mu / \mu$ constraints at z = 0.6874 and z = 0.88582 plotted
versus the scale factor $a = 1/(1+z)$.  The error bar at z = 0.6874 is $1\sigma$, however, the error bar 
at z = 0.88582 ($a=0.53$) includes systematic effects that increases the error to $\pm 10^{-7}$.  That
is the primary constraint utilized in this work.}  \label{fig-rad}
\end{figure}

The constraints on $\frac{\Delta \mu}{\mu}$ come from optical observations of redshifted electronic transitions of
molecular hydrogen and from radio observations of methanol and ammonia molecules.  The majority of the H$_2$ 
observations are at redshifts greater than two where the ultraviolet rest frame transitions enter the optical bands and the 
current radio observations are all at redshifts less than one.  The radio observations, however, are significantly more accurate 
than the optical H$_2$ observations.  The radio observations of methanol in PKS1830-211 \cite{kan15} $(\frac{\Delta \mu}
{\mu} = (-2.9 \pm10) \times 10^{-8})$ at a redshift  of 0.88582 is currently the tightest constraint on a variation of 
$\mu$.  In spite of the relatively low redshift the look back time is greater than half the age of the universe.  
Figure~\ref{fig-obs} and Table~\ref{tab-ob} show all of the constraints on the variation of $\mu$ with $1\sigma$ error 
bars.  The radio observations are shown separately in Fig.~\ref{fig-rad} since they are barely visible in Fig.~\ref{fig-obs}.

\begin{table}
\begin{tabular}{llllll}
\hline
Object & Redshift  & $\Delta \mu / \mu$ & $1\sigma$ error  & Ref.\\
\hline
J1443+2724 & 4.224 & $ -9.5 \times 10^{-6}$ & $\pm 7.6 \times 10^{-6}$ & \cite{bag15}\\
Q0347-383 & $3.0249$ & $2.1 \times 10^{-6}$ & $\pm 6. \times 10^{-6}$  & \cite{wen08}\\
Q0528-250 & $2.811$ & $3.0 \times 10^{-7}$ & $\pm 3.7 \times 10^{-6}$  &  \cite{kin11}\\ 
Q J0643-5041 & $2.659$ & $7.4 \times 10^{-6}$ & $\pm 6.7 \times 10^{-6}$ & \cite{vas14}\\
Q0405-443 & $2.5974$ & $10.1 \times 10^{-6}$ & $\pm 6.2 \times 10^{-6}$ &  \cite{kin08}\\
Q2348-011 & $2.426$ & $-6.8 \times 10^{-6}$ & $\pm 27.8 \times 10^{-6}$ & \cite{bag12}\\
He0027-1836 & $2.402$ & $-7.6 \times 10^{-6}$ & $\pm 1.0 \times 10^{-5}$ &  \cite{rah13}\\
Q01232+082 & $2.34$ & $1.9 \times 10^{-5}$ & $\pm 1.0 \times 10^{-5}$ & \cite{dap16}\\
J2123-005 & $2.059$ & $5.6 \times 10^{-6}$ & $\pm 6.2 \times 10^{-6}$ &  \cite{mal10}\\
PKS1830-211 & $0.88582$ & $-2.9 \times 10^{-8}$ & $\pm 5.7 \times 10^{-8}$ &  \cite{kan15}\\
B0218+357 & $0.6847$ & $-3.5 \times 10^{-7}$ & $\pm 1.2 \times 10^{-7}$ &  \cite{kan11}\\
\hline
\end{tabular}
\caption{Current best observational constraints on $\frac{\Delta \mu}{\mu}$}.  \label{tab-ob}
\end{table}
\subsection{$\alpha$ constraints} \label{ss-alphacon}
Contrary to the case with $\mu$ there are several thousand measurement of the fine structure splitting
at many redshifts of which several hundred are appropriate for testing for a variation of $\alpha$.
In two cases \cite{web01} and \cite{web11} temporal changes in $\alpha$ are reported with \cite{web11}
reporting both spatial and temporal changes at the $10^{-5}$ level. (See the contributions by Webb et al. in
these proceedings).  Subsequent work by \cite{mur16}, however, sees no variation at the $\frac{\Delta 
\alpha}{\alpha} = (0.4 \pm 1.7) \times 10^{-6}$ at the $1\sigma$ level where the error is the rms of the 
statistical and systematic errors. This is a significantly lower constraint than the reported variation by 
\cite{web11} of $\approx (-6.4 \pm 1.2) \times 10^{-6}$ and consistent with no change in $\alpha$.  
\cite{mur16} attributes the difference to known wavelength calibration errors in the previous analysis
in \cite{web01} and \cite{web11}.  For the purposes of this work the limits on a variation of $\alpha$
discussed in \cite{mur16} is taken as the primary limit.

\section{The Dependence of Fundamental Constants on the Physics Parameters} \label{s-funpar}
The numerical values of both $\alpha$ and $\mu$ depend on the values of the physics parameters
$\Lambda_{QCD}$, $\nu$ and $h$.  The following follows the discussion of \cite{coc07} done for
a different purpose but relevant to the current analysis.  The connection between the fundamental
constants and the physics parameters is probably most obvious for the proton to electron mass
ratio $\mu$ where the physics parameters set the mass of the proton and electron.  

\subsection{The proton to electron mass ratio} \label{ss-mu}
The fractional 
change of $\mu$, $\frac{d \mu}{\mu}$ by simple mathematics is
\begin{equation} \label{eq-dmu}
\frac{d\mu}{\mu} = \frac{dm_p}{m_p} -\frac{dm_e}{m_e}.
\end{equation}
The fractional change of the electron mass is easy since it is a fundamental particle whose mass is
set by the Higgs VEV and the electron Yukawa coupling such that $m_e=h_e\nu$ therefore
\begin{equation} \label{eq-dme}
\frac{dm_e}{m_e}=\frac{dh_e}{h_e} + \frac{d\nu}{\nu}.
\end{equation}
The mass of the proton, however, is much more complicated since it is a composite particle but the
fractional change is easier since the ratio eliminates some of the terms.  In \cite{coc07} the fractional
change of the proton mass is given by
\begin{equation} \label{eq-dmp}
\frac{dm_p}{m_p} = a \frac{d\Lambda_{QCD}}{\Lambda_{QCD}} + b (\frac{dh}{h}+\frac{d\nu}{\nu}).
\end{equation}
Both $a$ and $b$ are scalars of order unity whose sum by dimensional requirements should equal one
to ensure that the proton has units of mass.  Combining (\ref{eq-dme}) and (\ref{eq-dmp}) gives
\begin{equation} \label{eq-dmua}
\frac{d\mu}{\mu} = a \frac{d\Lambda_{QCD}}{\Lambda_{QCD}} + (b-1) (\frac{dh}{h}+\frac{d\nu}{\nu}).
\end{equation}
Here the common assumption that although the Yukawa couplings have different values their fractional changes
$\frac{d h}{h}$ should be the same is employed.  Next using $(a+b)=1$ $b$ is eliminated and the fractional
change of $\mu$ is
\begin{equation} \label{eq-qhy}
\frac{d\mu}{\mu} = a[\frac{d\Lambda_{QCD}}{\Lambda_{QCD}} - (\frac{dh}{h}+\frac{d\nu}{\nu})].
\end{equation}

\subsection{The fine structure constant} \label{ss-alpha}
Without additional information (\ref{eq-qhy}) only constrains the combination of $\Lambda_{QCD}$, $\nu$,
and $h$ rather than any of them individually.  The observational constraints on $\frac{d \alpha}{\alpha}$ can
provide some of that information since it constrains a different combination of the three parameters.  From
\cite{coc07} $\frac{d \alpha}{\alpha}$ depends on the three parameters as
\begin{equation} \label{eq-dal}
\frac{d\alpha}{\alpha} = R^{-1}[\frac{d\Lambda_{QCD}}{\Lambda_{QCD}} -\frac{2}{9}(\frac{dh}{h}+\frac{d\nu}{\nu})]
\end{equation}.
A new model dependent parameter $R$ is introduced that can range between unity and in excess of 100.  In
\cite{coc07} $R$ is taken as 36 based on unification arguments.  For the purposes of this work that value is accepted
partially since it is in the midrange of currently acceptable values.  Both (\ref{eq-qhy}) and (\ref{eq-dal}) effectively
depend on two variables, $\frac{d \Lambda_{QCD}}{\Lambda_{QCD}}$ and $(\frac{dh}{h} + \frac{d \nu}{\nu})$ which
provides a mechanism for solving for $\frac{d \Lambda_{QCD}}{\Lambda_{QCD}}$ as a function of the constraints
on the variance of $\mu$ and $\alpha$ and the model dependent parameters.

\subsubsection{The physics of $R$} \label{sss-R}
Various authors use different models and assumptions to set the value of $R$.  An example is \cite{din03}
who assumes that the variation in $\alpha$ is produced by temporal changes of the GUT unification scale $M_U$
which also makes the physics parameters time variable.  In this model $R$ is given by
\begin{equation} \label{eq-R}
R=\frac{2 \pi}{ 9 \alpha}\frac{\Delta b_3}{\frac{5}{3}\Delta b_1+\Delta b_2}
\end{equation}
where the $b_i$  are the beta function coefficients that scale $Q =\beta_i M < M_U$ .
At the unification scale $M_U$ all of the beta functions are unified to $b_U$.
$\Delta b_i$ is defined as $\Delta b_i  \equiv b_U-b_i$. The different gauge couplings 
$\alpha_i(Q)$ $(i=1,2,3)$ are then given by
\begin{equation} \label{eq-cou}
(\alpha_i(Q))^{-1} = (\alpha_U(M_U))^{-1}-\frac{b_i}{2\pi} ln(\frac{Q}{M_U})
\end{equation}
The GUT scale $M_U$ is allowed to change but $\alpha_U(M_{Pl})$ and
$M_{Pl}$ are held constant where $M_{Pl}$ is the Planck mass.
At the unification scale $R$ is given by
\begin{equation} \label{eq-bu}
R=\frac{2 \pi}{ 9 \alpha}\frac{b_U+3}{\frac{8}{3}b_U - 12}
\end{equation}
As $b_U$ becomes either positively or negatively large the value of $R$ approaches 36, the value used in \cite{coc07}.

\section{Observational Constraints on $\frac{d \Lambda_{QCD}}{\Lambda_{QCD}}$} \label{s-qcdcon}
The combination of constraints on the fractional variation of two fundamental constants $\mu$ and $\alpha$
provides the opportunity to put constraints on the fractional variation of $\Lambda_{QCD}$.  Eliminating
$(\frac{dh}{h} + \frac{d \nu}{\nu})$ from~\ref{eq-qhy} and~\ref{eq-dal} yields
\begin{equation} \label{eq-dqcdn}
\frac{d \Lambda_{QCD}}{\Lambda_{QCD}} = \frac{d \alpha}{\alpha}\frac{(b-1)R}{[(b-1)-\frac{2}{9}a]}
+ \frac{d \mu}{\mu}\frac{2}{9[(b-1)-\frac{2}{9}a]}
\end{equation}
where both factors $a$ and $b$ from (\ref{eq-dmp}) have been retained.  In \cite{coc07} $a=0.76$ and
$b=0.24$.  Again invoking $(a+b)=1$ (\ref{eq-dqcdn}) simplifies to
\begin{equation} \label{eq-dqcdab}
\frac{d \Lambda_{QCD}}{\Lambda_{QCD}} =\frac{9 R}{7}\frac{d \alpha}{\alpha} - \frac{2}{7 a}\frac{d \mu}{\mu}.
\end{equation}
Eliminating $\frac{d \Lambda_{QCD}}{\Lambda_{QCD}}$ provides a constraint on $(\frac{dh}{h} + \frac{d \nu}{\nu})$
of
\begin{equation} \label{eq-hvlim}
(\frac{dh}{h}+\frac{d\nu}{\nu}) = (\frac{9}{7})[R\frac{d \alpha}{\alpha}-\frac{1}{a}\frac{d \mu}{\mu}].
\end{equation}
Note that the leading terms on the right hand side of (\ref{eq-dqcdab}) and (\ref{eq-hvlim}) are identical.

\subsection{Individual constraints on $\frac{\Delta \nu}{\nu}$ and $\frac{ \Delta h}{h}$} \label{ss-hy}
At the expense of additional model dependence it is possible to individually constrain $\frac{d \nu}{\nu}$
and $\frac{d h}{h}$.  In the standard model there is an established relationship between the fractional variation
of the Higgs VEV $\nu$ and the fractional variation of the Yukawa couplings $h$ given by \cite{coc07}.  
\begin{equation} \label{eq-s}
\frac{d \nu}{\nu} = S\frac{d h}{h}.
\end{equation}
The value of $S$ is model dependent so the constraints on  $\frac{d \nu}{\nu}$ and $\frac{d h}{h}$ are
doubly model dependent.  These constraints come from using (\ref{eq-s}) in (\ref{eq-hvlim}). The $\pm$ in the
 $\frac{\Delta \alpha}{\alpha}$ and $\frac{\Delta \mu}{\mu}$
terms in equations~\ref{eq-dnu} and~\ref{eq-h} simply reflect that limits on $\frac{d \alpha}{\alpha}$ and
$\frac{ d \mu}{\mu}$ are plus or minus the quoted error.  As stated in section~\ref{s-mod} the terms are combined
as a root mean square when evaluated.
\begin{equation} \label{eq-dnu}
\frac{\Delta \nu}{\nu} =\frac{9}{7} \frac{S}{(1+S)}[(R\frac{\pm \Delta \alpha}{\alpha}) - \frac{1}{a}(\frac{\pm \Delta \mu}{\mu})].
\end{equation}
Similarly the limit on $\Delta h/h$ is
\begin{equation} \label{eq-h}
\frac{\Delta h}{h} =  \frac{9}{7}\frac{1}{(1+S)}[(R\frac{\pm \Delta \alpha}{\alpha}) - \frac{1}{a}(\frac{\pm \Delta \mu}{\mu})].
\end{equation}

\section{Constraints From a Given Model} \label{s-mod}
Since the discussion has followed the work of \cite{coc07} the model used in that reference is used as an example
for producing constraints on the physics parameters.  It is also a constraint on the particular model as well.  The
values for the model dependent parameters in \cite{coc07} are $a=0.76$, $b=0.24$ (which add to one), $R=36$
and $S=160$. Note that with this parameter set and the observational constraints the common leading term, the
$\alpha$ term, in (\ref{eq-dqcdab}) and (\ref{eq-hvlim}) dominate the constraints on the parameters.  The 
constraints are
\begin{equation} \label{eq-qcdc}
\Delta \Lambda_{QCD} /  \Lambda_{QCD} \leq \pm 7.9 \times 10^{-5}
\end{equation}
\begin{equation}
\Delta \nu / \nu \leq \pm 7.9 \times 10^{-5}
\end{equation}
\begin{equation}
\Delta h / h \leq \pm 4.9 \times 10^{-7}.
\end{equation} 
The  look back time for the constraints is the average look back time of the $\alpha$ observations at a
redshift of 1.54 equal to 9.4 gigayears or roughly $70\%$ of the age of the universe.

\begin{figure}
  \vspace{30pt}
\scalebox{.7}{\includegraphics{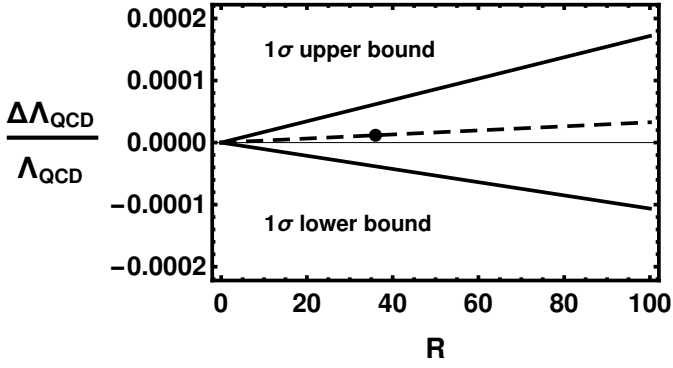}}
  \caption{The figure indicates the 1$\sigma$ variation of the limit on $d \Lambda_{QCD} / \Lambda_{QCD}$ 
as a function of the model parameter $R$. The dashed line indicates the limit on $d \Lambda_{QCD} / 
\Lambda_{QCD}$ if the measured values of $d \alpha / \alpha$ and $d \mu/\mu$ are used
rather than the limits.  The dot is at $R=36$ which is the example value.  Note that although it is not apparent at
the scale of the figure the limit on $d \Lambda_{QCD} / \Lambda_{QCD}$ at $R=0$ is not zero but rather the small
$ \frac{2}{11 a}\frac{d \mu}{\mu}$ term in (\ref{eq-dqcdab}) that does not depend on $R$.} 
\label{fig-qcdr}
\end{figure}

The observational constraints are evaluated via rms summing of the $1\sigma$ $\alpha$ and $\mu$ terms for 
each parameter.  This assumes that the $1\sigma$ errors are centered on zero rather than the "measured"
values of the observations. It was noted earlier that the $R$ model parameter has a range of model values.  
Figure~\ref{fig-qcdr} shows the variation of the limit on $\Delta \Lambda_{QCD} / \Lambda_{QCD}$ as a 
function of $R$.

\section{A Model Dependent Limit on $\frac{\Delta \alpha}{\alpha}$} \label{s-moda}
The observational constraints on the physics parameters are dominated by the $\alpha$ term both
because of its less stringent observational limit and by the larger coefficient in the model of \cite{coc07}.
That model, however, predicts a relationship between a variation of $\alpha$ and a variation of $\mu$
given by
\begin{equation} \label{eq-dma}
\frac{d \alpha}{\alpha} = \frac{1}{R} \frac{d \mu}{\mu}.
\end{equation}
The model therefore predicts the fractional variation of $\alpha$ should be smaller than the variation
of $\mu$ by a factor of 1/36.  It is of some interest to see how the constraints on the physics parameters
change if this model dependent limit of  $\frac{\Delta \alpha}{\alpha} \leq \pm2.8 \times 10^{-9}$ on the 
fractional change of $\alpha$ is imposed.  The new constraint on $\frac{d \Lambda_{QCD}}{\Lambda_{QCD}}$
is
\begin{equation} \label{eq-dqcdmod}
\frac{\Delta \Lambda_{QCD}}{\Lambda_{QCD}} \leq [\pm (2.8 \times 10^{-9})\frac{9 R}{7} \pm 10^{-7}\frac{2}{7 a}]
\leq \pm 1.7 \times 10^{-7}.
\end{equation}
Similar replacements in (\ref{eq-dnu}) and (\ref{eq-h}) yield constraints on the time variation of $\nu$ and $h$ of
\begin{equation} \label{eq-vmod}
\frac{\Delta \nu}{\nu} \leq (\frac{9}{7}) (\frac{S}{S+1})[\pm R(2.8 \times 10^{-9}) \pm\frac{10^{-7}}{a}]
\leq \pm 2.1 \times 10^{-7}
\end{equation}
\begin{equation} \label{eq-hmod}
\frac{\Delta h}{h} \leq (\frac{9}{7}) (\frac{1}{S+1})[\pm R(2.8 \times 10^{-9}) \pm\frac{10^{-7}}{a}]
\leq \pm 1.3 \times 10^{-9}.
\end{equation}
The net result of the model dependent limit on $\frac{d \alpha}{\alpha}$ is a very stringent set of limits on the
fractional change of the parameters at the look back time of the $\mu$ constraints which is greater than half the
age of the universe.  This constraint severely limits the parameter space of theories that predict significant fractional 
changes in $\Lambda_{QCD}$, $\nu$ or $h$

\section{Conclusions}
It is difficult to test for time variability of the primary particle physics parameters such as the Quantum 
Chromodynamic Scale, the Higgs Vacuum Expectation Value and the Yukawa couplings.  It is, however,
relatively easy to use spectra of objects in the early universe to test for time variability of dimensionless
fundamental constants whose numerical values depend on the physics parameters.  Individually the constants
only constrain a combination of the parameters but combining the observational constraints on the variability
of two or more constants provides model dependent constraints on the fractional variability of the QCD scale and
a combination of the fractional variability of the Higgs VEV and the Yukawa couplings.  Introduction of an additional
model dependent parameter sets limits on the fractional variability of the Higgs VEV and the Yukawa couplings
separately.

The constraints on the fractional time variability of the physics parameters limits the parameter space of new
physics theories that require a time variability of any of the three basic physics parameters.  It is recommended
that observed constraints on the variability of dimensionless fundamental constants become another important
tool in evaluating the validity of non-standard physics and cosmology theories.


\section{References}
\begin{thebibliography}{999}
\bibitem{web01} Webb, J.K., King, Murphy, M.T., Flambaum, V.V., Dzuba, V.A., Barrow, J.D., Churchill, C.W., 
	Prochaska, J.X. \& Wolfe, A.M. Further Evidence for Cosmological Evolution of the Fine Structure Constant.
	{\em PRL}, {\bf 2001}, {\em 87}, 091301-1-4.
\bibitem{cal02}	Calmet, X. and Fritzsch, H. Symmetry Breaking and Time Variation of Gauge Couplings.
	{\em Phys. Lett. B}, {\bf 2002}, {\em 540}, 173-178.
\bibitem{cam95} Campbell, B.A. and Olive, K.A. Nucleosynthesis and the time dependence of fundamental couplings, 
	{\em Phys. Lett. B}, {\bf 1995}, {\em 345}, 429-434
\bibitem{cha07} Chamoun, N., Landou, S.J., Mosquera, M.E. and Vucetich, H. Helium and deuterium abundances as 
	a test for the time variation of the fine structure constant and the Higgs vacuum expectation value. {\em J. Phys. G: 
	Nucl. Part. Phys.}, {\bf2007}, {\em 34}, 163-176.
\bibitem{coc07} Coc, A., Nunes, N.J., Olive, K.A., Uzan, J-P, \& Vangioni, E. Coupled variations of fundamental couplings 
	and primordial nucleosynthesis, {\em Phys. Rev. D}, {\bf2007},  {\em 76}, 023511 1-12.
\bibitem{den08} Dent, T. Fundamental constants and their variability in theories of High Energy Physics. {\em Eur. Phys. J. 
	Special Topics}. {\bf 2008}, {\em 163}, 297–313. 
\bibitem{din03} Dine, M., Nir, Y., Raz, G. and Volansky, T. Time variations in the scale of grand unification. {\em Phys. Rev. D}, 
	{\bf 2003}, {\em 67}, 015009 1-6.
\bibitem{lan02} Langacker, P., Segre, G. and Strassler, M.J. Implications of gauge unification for time variation of the 
	fine structure constant. {\em Phys. Lett. B}, {\bf 2002}, {\em 528}, 121-128 
\bibitem{lan04} Langacker, P. Time Variation of Fundamental Constants as a Probe of New Physics. {\em Int. Jr. Mod. Phys. A}, 
	{\bf 2004}, {\em 19}, 157-165.
\bibitem{uza11} Uzan, J-P Varying constants, Gravitation and Cosmology. {\em Living Rev. Relativity}, {\bf 2011}, {\em 14} 
	2-155.
\bibitem{thm75} Thompson, R.I. The Determination of the Electron to Proton Inertial Mass Ratio via Molecular Transitions. 
	{\em Astrophys. Lett.}, {\bf 1975}, {\em15}, 3-4.
\bibitem{kan15} Kanekar, N., Ubachs, W., Menten, K.M., Bagdonaite, J., Brunthaler, A., Henkel, Muller, C.S., Bethlem, H.L. 
		and Dapra, M. Constraints on changes in the proton-electron mass ratio using methanol lines. {\em MNRAS}, {\bf 2015}, 
	448, L104-L108.
\bibitem{bag15} Bagdonaite, J., Ubachs, W., Murphy, M.T. \& Whitmore, J. B. Constraint on a Varying Proton-Electron Mass 
	Ratio 1.5 Billion Years after the Big Bang. {\em PRL},{\bf 2015}, {em 114}, 071301 1-6.
\bibitem{wen08} Wendt, M. and Reimers, D. Variability of the proton-to-electron mass ratio on cosmological scales. 
	{\em Eur. Phys. J. Special Topics} {\bf 2008}, {\em 163} 197–206.
\bibitem{kin11} King, J.A., Murphy, M.T., Ubachs, W. and Webb, J.K. New constraint on cosmological variation of the 
	proton-to-electron mass ratio from Q0528-250. {\em MNRAS}, {\bf 2011}, {em 417}, 3010–3024.
\bibitem{vas14} Vasquez, F. A., Rahamani, H., Noterdaeme, N., Petitjean, P., Srianand, R. and Ledoux, C. Molecular hydrogen 
	in the zabs = 2:66 damped Lyman-alpha absorber towards Q J 0643-5041. {\em A\&A}, {\bf 2014}, {\em 562}, A88 1-21.
\bibitem{kin08} King, J. A., Webb, J.K., Murphy, M.T., and Carswell, R.F. Stringent Null Constraint on Cosmological Evolution 
	of the Proton-to-Electron Mass Ratio. {\em PRL}, {\bf 2008}, {\em 101}, 251304 1-4.
\bibitem{bag12} Bagdonaite, J., Murphy, M.T., Kaper, L. and Ubachs, W. Constraint on a variation of the proton-to-electron 
	mass ratio from H$_2$ absorption towards quasar Q2348-011. {\em MNRAS}, {\bf 2012}, {\em 421}, 419-425.
\bibitem{rah13} Rahmani, H. et al. The UVES large program for testing fundamental physics – II. Constraints on a change 
	in $\mu$ towards quasar HE 0027-1836. {\em MNRAS}, {\bf 2013}, {\em 435}, 861–878.
\bibitem{dap16}  Dapra,M., van der Laan, M., Murphy, M.T. and Ubachs, W. Constraint on a varying proton-to-electron mass 
	ratio from H2 and HD absorption at zabs = 2.34. {\em arXiv:1611.05191v2 [astro-ph.CO]} {\bf 2016} 
\bibitem{mal10} Malec, A.L. et al. Keck telescope constraint on cosmological variation of the proton-to-electron mass ratio.
	{\em MNRAS}, {\bf 2010}, {em 403}, 1541-1555.
\bibitem{kan11}, Kanekar, N. Constraining Changes in the Proton–Electron Mass Ratio with Inversiona and Rotational Lines.
	{\em ApJL}, {\bf 2011}, {\em 728} L12, 1-5.
\bibitem{web11} Webb, J.K., J.A., Murphy, M.T., Flambaum, V.V., Carswell, R.F., \& Bainbridge, M.B. Indications of a 
	spatial variation of the fine structure constant. {\em PRL}, {\bf 2011}, {\em 107}, 191101-1-5.
\bibitem{mur16} Murphy, M.T., Malec, A. and Prochaska, J.X. Precise limits on cosmological variability of the fine-structure 
	constant with zinc and chromium quasar absorption lines. {\em MNRAS}, {\bf 2016}, {\em 461}, 2461-2479.
\end{thebibliography}

\end{document}